\documentclass[pra, showpacs]{revtex4}
\usepackage{amssymb}
\usepackage{bm}
\usepackage{graphicx}%
\begin{document}
%\draft
\title
{Drag of superfluid current in bilayer Bose systems}
\author{D.\,V.\,Fil}
\email{fil@isc.kharkov.com} \affiliation{ Institute for Single
Crystals, National Academy of Sciences of Ukraine, Lenin av. 60,
Kharkov 61001, Ukraine}
\author{S.\,I.\,Shevchenko}
\email{shevchenko@ilt.kharkov.ua}
\affiliation{%
B.\,I.\, Verkin Institute for Low Temperature Physics and
Engineering National Academy of Sciences of Ukraine, Lenin av. 47
Kharkov 61103, Ukraine}
%\date{\today}

\begin{abstract}
An effect of  nondissipative drag of  a superfluid flow in a
system of two Bose gases confined in two parallel quasi
two-dimensional traps is studied. Using an approach based on
introduction of density and phase operators we compute the drag
current at zero and finite temperatures for arbitrary ratio of
densities of the particles in the adjacent layers. We demonstrate
that in a system of two ring-shape traps the "drag force"
influences on the drag trap in the same way as an external
magnetic flux influences on a superconducting ring. It allows to
use the drag effect to control persistent current states in
superfluids and opens a possibility for implementing a Bose analog
of the superconducting Josephson flux qubit.
\end{abstract}


\maketitle

\section{INTRODUCTION}
 The existence of nondissipative supercurrents is a common
feature of superconducting and superfluid systems. Among various
applications of superconductivity  considerable attention was
given to the use of superconducting circuits as very sensitive
magnetometers (superconducting quantum interferometer devices). At
present an interest to such systems is renewed in view of a
possibility to use superconducting circuits with weak links as
elements of quantum computers (Josephson qubits). Basing on a
similarity between superfluids and superconductors one can expect
that the former ones may also be used for implementing qubits.

Supercurrent in superconductors is coupled to vector potential of
electromagnetic fields. It allows to control persistent current
states by external fields. Obviously, there is no such a channel
for a control in uncharged superfluid Bose systems. In this paper
we study another possibility  based on a nondissipative drag
effect.

The drag in normal systems has been investigated experimentally
and theoretically by many authors (see, for instance, reviews
\cite{1,2}). The main attention was given to the study of bilayer
electron  systems in semiconductor heterostructures. In such
systems an interlayer drag effect takes place. The effect is
caused by electron-electron scattering processes and it reveals
itself in an appearance of a drag voltage in one layer when a
normal current flows in the adjacent layer. If the former layer is
in a closed circuit, the drag voltage induces a drag current
flowing through the circuit. The effect is accompanied by a
dissipation of energy and takes place only at finite temperatures.
Roughly, the drag voltage increases by the $T^2$-law (deviation
from this law, observed experimentally \cite{3}, is connected with
a phonon contribution to the interaction between the carriers
\cite{4}).

In superfluid and superconducting systems another  kind of drag
may take place. This drag is nondissipative % in its nature
and is connected with a redistribution of a supercurrent between
two superfluid (superconducting) components. In difference with
the drag in a normal state the superfluid drag has the largest
value at zero temperatures and  decreases under increasing the
temperature. On the existence of nondissipative drag in superfluid
systems was pointed out for the first time in the paper by Andreev
and Bashkin \cite{5}. In this paper a three-velocity hydrodynamic
model of a $^3$He-$^4$He mixture was developed. It has been shown
that superfluid behavior of such  systems can be described under
accounting the "drag"\ term in the free energy. This term is
proportional to the scalar product of  superfluid velocities of
two components times the difference between the effective and the
bare masses of $^3$He atoms. In the paper by Duan and Yip \cite{6}
the nondissipative drag effect in superconductors has been
studied. The authors of \cite{6} argue that the value of drag can
be obtained from the energy of zero-point fluctuations. It was
shown that  this energy contains a "drag" term analogous to one
obtained in \cite{5} in the hydrodynamic approach. The theory of
the nondissipative drag in  a bilayer system of charged bosons was
developed by Tanatar and Das \cite{7} and  by Terentjev and
Shevchenko \cite{8}.

The existence of nondissipative drag in a system of two
one-dimensional wires in a persistent current state was predicted
by Rojo and Mahan \cite{9} . It was also shown by Duan \cite{10}
that nondissipative drag is responsible for an emergence of an
interlayer Hall voltage in bilayer electron systems in the
fractional quantum Hall regime.

Basing on the previous studies one can consider  nondissitive drag
as a fundamental property of  systems with  macroscopic quantum
coherence. For the system of uncharged bosons this effect is
especially important, since the "drag force" plays the role
similar to the role of the vector potential of  magnetic field in
superconductors. It opens new possibilities to observe the effects
caused by  phase coherence in such systems. One of the goals of
this paper is  to point out on this analogy. In particular, we
show that the nondissipative drag effect allows to realize an
entangled flux state in  a superfluid ring with  a weak Josephson
link.

Now a great attention is given to the study of ultracold
alkali-metal vapours confined in magnetic and optical traps, where
Bose-Einstein condensation of atoms was observed \cite{11}.
Advances in technology allow to manipulate  parameters of such
systems  and make ultracold atomic gases an unique object for the
study of various quantum mechanical phenomena.

In this paper we study the effect of nondissipative drag in a
system of two quasi two-dimensional atomic Bose gases confined in
two parallel traps. To describe such a situation we take into
account that  densities of  atoms in the drive and  the drag
layers can  be non-equal. In previous studies \cite{7,8} only the
case of two layers with equal densities of the particles was
considered. Another important factor is the temperature. In atomic
gases it is of order or higher than the energy of intralayer
interactions.  Previously, the dependence of nondissipative drag
on the temperature was treated only qualitatively \cite{6,8}. Here
we  study the temperature dependence quantitatively. We also
evaluate the value of the drag for concrete mechanisms of
interlayer interaction in atomic Bose gases.

\section{Model and approach}
\label{s2}

  The geometry of the Bose cloud can be modified
significantly under a variation of a configuration of external
fields forming a trap. When the confining potential is strongly
anisotropic and the temperature as well as the chemical potential
are smaller than the separation between the energy levels of
spatial quantization in one direction  the Bose gas  can be
treated as a two-dimensional one. Recently, low-dimensional atomic
gases were realized experimentally \cite{12}.

Bose clouds of a ring shape can be created by using toroidal
traps. A configuration of two toroidal traps situated one above
another is convenient for the study of the drag effect. As it
follows from the further consideration if one excites a
circulating superflow in one trap it inevitably leads to a
redistribution of this superflow between two traps and superfluid
currents appear in both rings.

Main features of the drag effect can be understood from the study
of a system of two uniform two-dimensional Bose gases situated in
parallel layers. The Hamiltonian of the system can be presented in
the form
\begin{equation}\label{1}
  H=\sum_{l=1,2} (E_l - \mu_l N_l)+\frac{1}{2}\sum_{l,l'=1,2}
  E_{ll'}^{int},
\end{equation}
where
\begin{equation}\label{2}
  E_l=\int d^2 r \frac{\hbar^2}{2 m}
  [\nabla\hat{\Psi}^+_l({\bf r})]\nabla\hat{\Psi}_l({\bf r})
 \end{equation}
is the kinetic energy,
\begin{equation}\label{3}
  E_{l l'}^{int}=\int d^2  r  d^2 r^\prime
 \hat{\Psi}^+_l({\bf r})\hat{\Psi}^+_{l'}({\bf r}')V_{l l'}({\bf r}-{\bf r}')
 \hat{\Psi}_{l'}({\bf r}')
 \hat{\Psi}_{l}({\bf r})
 \end{equation}
is the energy of the intralayer ($l=l'$) and interlayer ($l\ne
l'$) interaction, and $N_l=\int d^2 r \hat{\Psi}^+_l({\bf
r})\hat{\Psi}_l({\bf r})$. Here $\hat{\Psi}$ is the Bose field
operator, $l$, the layer index, ${\bf r}$, the two-dimensional
radius-vector lying in the layer, and $\mu_l$, the chemical
potentials. To be more specific we consider the case of point
interaction between the atoms: $V_{1 1}({\bf r})=V_{2 2}({\bf
r})=\gamma \delta({\bf r})$, $V_{1 2}({\bf r})=V_{2 1}({\bf
r})=\gamma' \delta({\bf r})$ with $\gamma>0$ and
$|\gamma'|<\gamma$. Assuming the barrier between two traps is
quite high we neglect the tunneling between the layers.

For further analysis it is convenient to use the density and phase
operator approach (see, for instance, \cite{13,14}). The approach
is based on the following representation for the Bose field
operators
\begin{equation}\label{4}
  \hat{\Psi}_l({\bf r})=\exp\left[i\varphi_{l}({\bf r})+i\hat{\varphi}_l({\bf r})\right]
  \sqrt{\rho_l+\hat{\rho}_l({\bf r})},
\end{equation}
\begin{equation}\label{5}
  \hat{\Psi}_l^+({\bf r})=\sqrt{\rho_l+\hat{\rho}_l({\bf r})}
  \exp\left[-i\varphi_{l}({\bf r})-i\hat{\varphi}_l({\bf
  r})\right],
\end{equation}
where $\hat{\rho}_l$ and $\hat{\varphi}_l$ are the density and
phase fluctuation operators,  $\rho_l=\langle\hat{\Psi}_l^+({\bf
r})\hat{\Psi}_l({\bf r})\rangle$ is the $c$-number term of the
density operator (one can see that it is just the density of atoms
in the layer $l$), $\varphi_{l}({\bf r})$ is the $c$-number term
of the phase operator (in the approach considered the inclusion of
this term in the phase operator allows to describe the states with
nonzero average superflows).

Substituting Eqs.  (\ref{4}), (\ref{5}) into Hamiltonian (\ref{1})
and expanding it in series in powers of $\hat{\rho}_l$ and
$\nabla\hat{\varphi}_l$ we arrive to the expression
\begin{equation}\label{6}
  H=H_{0}+H_{1}+H_{2}+\ldots
\end{equation}
In (\ref{6}) the term
\begin{equation}\label{7}
  H_0=\int d^2 r\left\{\sum_l \left[\frac{\hbar^2}{2 m}\rho_l
  \Big(\nabla\varphi_l({\bf r})\Big)^2
  +\frac{\gamma}{2}\rho_l^2-\mu_l\rho_l\right]+\gamma'\rho_1\rho_2\right\}
\end{equation}
does not contain the operator part, the term
\begin{equation}\label{8}
  H_1=\int d^2 r\left(\sum_l\left\{\left[\frac{\hbar^2}{2m}
  \Big(\nabla\varphi_l({\bf r})\Big)^2+\gamma\rho_l+\gamma'\rho_{3-l}-
  \mu_l\right]\hat{\rho}_l({\bf r})
  +\frac{\hbar^2}{m}\rho_l[\nabla\varphi_l({\bf r})]\nabla\hat{\varphi}_l({\bf r})
  \right\}\right)
\end{equation}
is linear in the phase and density fluctuation operators, and the
term
\begin{eqnarray}\label{9}
  H_2=\int d^2 r \Bigg(\sum_l\frac{\hbar^2}{2
  m}\Bigg[\frac{\Big(\nabla\hat{\rho}_l({\bf r})\Big)^2}{4\rho_l}
  +\rho_l\Big(\nabla\hat{\varphi}_l({\bf r})\Big)^2+
  [\nabla\varphi_l({\bf r})]\Big(\hat{\rho}_l({\bf r})\nabla\hat{\varphi}_l({\bf r})
  +[\nabla\hat{\varphi}_l({\bf r})]\hat{\rho}_l({\bf r})\Big)+\cr
  \frac{i}{2}\Big([\nabla\hat{\rho}_l({\bf r})]\nabla\hat{\varphi}_l({\bf r})-
  [\nabla\hat{\varphi}_l({\bf r})]\nabla\hat{\rho}_l({\bf r})\Big)\Bigg]
  +\frac{\gamma}{2}\Big[\Big(\hat{\rho}_1({\bf r})\Big)^2+
  \Big(\hat{\rho}_2({\bf r})\Big)^2\Big]
  +\gamma'\hat{\rho}_1({\bf r})\hat{\rho}_2({\bf r})\Bigg)
\end{eqnarray}
is quadratic in $\nabla\hat{\varphi}_l$ and $\hat{\rho}_l$
operators.

If the chemical potentials are fixed  the Hamiltonian $H_0$ is
minimized under  conditions
\begin{equation}\label{10}
\frac{\hbar^2}{2m}\Big(\nabla\varphi_l({\bf
r})\Big)^2+\gamma\rho_l+\gamma'\rho_{3-l}-
  \mu_l=0,
\end{equation}
\begin{equation}\label{11}
  \nabla\Big(\rho_l\nabla\varphi_l({\bf r})\Big)=0
\end{equation}
Fulfillment of Eqs. (\ref{10}), (\ref{11}) means that the linear
in density and phase fluctuation operator term $H_1$ in the
Hamiltonian vanishes. One should note that, as it follows from Eq.
(\ref{10}), the densities of the components do not depend on the
coordinates only when the phase gradients remain space independent
as well.

The quadratic part of the Hamiltonian determines the spectrum of
elementary excitations. Hereafter we will neglect the higher order
terms in the Hamiltonian (\ref{6}). These terms describe the
scattering of the quasiparticles and they can be omitted if the
temperature is much smaller than the critical temperature
($T_c\sim \hbar^2 \rho/m$).

\section{Drag current}
\label{s3}

The operator of the density of the current
\begin{equation}\label{12}
  \hat{{\bf j}}_l=\frac{i\hbar}{2m}\left[(\nabla
  \hat{\Psi}_l^+)\hat{\Psi}_l -
  \hat{\Psi}_l^+\nabla\hat{\Psi}_l\right],
\end{equation}
rewritten in terms of the phase and density operators, has the
form
\begin{equation}\label{13}
  \hat{{\bf j}}_l=\frac{\hbar}{m}
  \sqrt{\rho_l+\hat{\rho}_l}[\nabla(\varphi_l+\hat{\varphi}_l)]
  \sqrt{\rho_l+\hat{\rho}_l}.
\end{equation}
Expanding (\ref{13}) in powers of the density and phase
fluctuation operators and neglecting the terms of order higher
than quadratic one we obtain the following expression for the mean
value of the density of the current
\begin{equation}\label{14}
{\bf j}_l=\frac{\hbar}{m}\rho_l\nabla\varphi_l+\frac{\hbar}{2 m}
\left(\langle[\nabla\hat{\varphi}_l]\hat{\rho}_l\rangle+
\langle\hat{\rho}_l\nabla\hat{\varphi}_l\rangle \right).
\end{equation}
To derive Eq. (\ref{14}) we take into account that
$\langle\hat{\varphi}_l\rangle=\langle\hat{\rho}_l\rangle=0$.

To compute the averages in (\ref{14}) we  rewrite the quadratic
part of the Hamiltonian in terms of the operators of creation and
annihilation of the elementary excitations. In the absence of the
interlayer interaction ($\gamma'=0$) it can be done by the
substitution
\begin{equation}\label{15}
  \hat{\rho}_l({\bf r})=\sqrt{\frac{\rho_l}{S}}\sum_{\bf k}
  e^{i{\bf k r}}\sqrt{\frac{\epsilon_{k}}{E_{l k}}}
  \left[ b_{l}({\bf k})+b^+_{l}(-{\bf
  k})\right],
\end{equation}
\begin{equation}\label{16}
  \hat{\varphi}_l({\bf r})=\frac{1}{2i}\sqrt{\frac{1}{\rho_l S}}\sum_{\bf k}
  e^{i{\bf k r}}\sqrt{\frac{E_{l k}}{\epsilon_{k}}}
  \left[ b_{l}({\bf k})-b^+_l(-{\bf
  k})\right],
\end{equation}
where the operators $b_l^+$, $b_l$ satisfy the Bose commutation
relations. Here $S$ is the area of the system, $\epsilon_k=\hbar^2
k^2/2 m$ is the spectrum of free atoms, and
$E_{lk}=\sqrt{\epsilon_{k}(\epsilon_{k}+2\gamma\rho_l)}$ is the
spectrum of  elementary excitations at $\gamma'=0$ and
$\nabla\varphi_l=0$.

In the case considered the substitution (\ref{15}), (\ref{16})
reduces the Hamiltonian (\ref{9}) to the form
\begin{equation}\label{17}
  H_2=\sum_{l{\bf k}} \left[{\cal E}_{l}({\bf k})  b^+_{l}({\bf k})  b_{l}({\bf
  k})+\frac{1}{2}(E_{l k}-\epsilon_k)\right]
  +\sum_{\bf k} g_k\left[b^+_{1}({\bf k})  b_{2}({\bf k})+b_{1}({\bf k})
   b_{2}({\bf
  -k})+h.c.\right],
\end{equation}
where
\begin{equation}\label{18}
  {\cal E}_{l}({\bf k})=E_{l k}+\frac{\hbar^2}{m}  {\bf
k} \nabla\varphi_l,
\end{equation}
\begin{equation}\label{19}
  g_k=\gamma' \epsilon_k \sqrt{\frac{\rho_1\rho_2}{E_{1 k}E_{2
k}}}.
\end{equation}
The Hamiltonian (\ref{17}) contains non-diagonal in Bose creation
and annihilation operator terms and  it can be diagonalized using
the u-v transformation
\begin{eqnarray}\label{20}
  b_{l}({\bf k})=u_{l\alpha}({\bf k})\alpha({\bf k})+u_{l\beta}({\bf k})\beta ({\bf
  k})+v_{l\alpha}({\bf k)}\alpha^+(-{\bf k})+v_{l\beta}({\bf k})\beta^+(-{\bf
  k})
\end{eqnarray}
(see [\cite{15}]) that reduces the Hamiltonian (\ref{17}) to the
form
\begin{equation}\label{21}
 H_2= \sum_{\bf k} \left[{\cal E}_{\alpha}({\bf k})  \left(\alpha^+({\bf k})  \alpha({\bf
  k})+\frac{1}{2}\right)+{\cal E}_{\beta}({\bf k})  \left(\beta^+({\bf k})  \beta({\bf
  k})+\frac{1}{2}\right)-\epsilon_k\right].
\end{equation}

It is convenient to present the $u-v$ coefficients and the
energies of the elementary excitations as  series in powers of
$g_k$. The $u-v$ coefficients read as
\begin{equation}\label{22}
  \left(\matrix{u_{1\alpha}({\bf k})&u_{1\beta}({\bf k})\cr
  u_{2\alpha}({\bf k})&u_{2\beta}({\bf k})}\right)=\left(\matrix{A_{\bf k}
     &  - \frac{ g_k}{{\cal E}_{1}({\bf k})-{\cal E}_{2}({\bf k})}\cr
      \frac{ g_k}{{\cal E}_{1}({\bf k})-{\cal E}_{2}({\bf k})}&B_{\bf
     k}}\right)+{\cal O}(g_k^3),
 \end{equation}
\begin{eqnarray}\label{23}
   \left(\matrix{v_{1\alpha}({\bf k})&v_{1\beta}({\bf k})\cr
  v_{2\alpha}({\bf k})&v_{2\beta}({\bf k})}\right)=\cr
  \left(\matrix{-\frac{ g_k^2 [{\cal E}_{2}({\bf
  k})+{\cal E}_{2}(-{\bf k})]}{[{\cal E}_{1}({\bf
  k})+{\cal E}_{1}(-{\bf k})][{\cal E}_{1}(-{\bf
  k})-{\cal E}_{2}(-{\bf k})][{\cal E}_{1}(-{\bf
  k})+{\cal E}_{2}({\bf k})]}
     &  - \frac{ g_k}{{\cal E}_{1}({\bf k})+{\cal E}_{2}(-{\bf k})}\cr
      -\frac{ g_k}{{\cal E}_{1}(-{\bf k})+{\cal E}_{2}({\bf k})}&
\frac{ g_k^2 [{\cal E}_{1}({\bf
  k})+{\cal E}_{1}(-{\bf k})]}{[{\cal E}_{2}({\bf
  k})+{\cal E}_{2}(-{\bf k})][{\cal E}_{1}(-{\bf
  k})-{\cal E}_{2}(-{\bf k})][{\cal E}_{1}({\bf
  k})+{\cal E}_{2}(-{\bf k})]}}
\right)+{\cal O}(g_k^3),
 \end{eqnarray}
where
\begin{equation}\label{24}
A_{\bf k}=1-\frac{g_k^2}{2}\left(\frac{1}{[{\cal E}_{1}({\bf
k})-{\cal E}_{2}({\bf k})]^2}-\frac{1}{[{\cal E}_{1}({\bf
k})+{\cal E}_{2}(-{\bf k})]^2}\right),
\end{equation}
\begin{equation}\label{25}
B_{\bf k}=1-\frac{g_k^2}{2}\left(\frac{1}{[{\cal E}_{1}({\bf
k})-{\cal E}_{2}({\bf k})]^2}-\frac{1}{[{\cal E}_{1}(-{\bf
k})+{\cal E}_{2}({\bf k})]^2}\right).
\end{equation}

The spectra of the elementary excitations are found to be
\begin{equation}\label{26}
{\cal E}_{\alpha}({\bf k})={\cal E}_{1}({\bf k})+
g_k^2\left[\frac{1} {{\cal E}_{1}({\bf
  k})-{\cal E}_{2}({\bf k})}-\frac{1}
{{\cal E}_{1}({\bf
  k})+{\cal E}_{2}(-{\bf k})}\right]+{\cal O}(g_k^4),
\end{equation}
\begin{equation}\label{27}
{\cal E}_{\beta}({\bf k})={\cal E}_{2}({\bf k})-
g_k^2\left[\frac{1} {{\cal E}_{1}({\bf
  k})-{\cal E}_{2}({\bf k})}+\frac{1}
{{\cal E}_{1}({-\bf
  k})+{\cal E}_{2}({\bf k})}\right]+{\cal O}(g_k^4).
\end{equation}

One can see that the small parameter of the expansion is
$g_k/|{\cal E}_{1}({\bf k})-{\cal E}_{2}({\bf k})|\ll 1$. The last
inequality takes place for all $k$, if
$\gamma'\max(\rho_1,\rho_2)\ll \gamma |\rho_1-\rho_2|$. Since in
most cases of interest the interlayer interaction is much smaller
than the intralayer one, for the bilayer systems with different
densities in the adjacent layers one can neglect the ${\cal
O}(g_k^3)$ and higher order terms in Eqs. (\ref{22}), (\ref{23}),
(\ref{26}), (\ref{27}).

Using representation (\ref{15}), (\ref{16}) we obtain from
(\ref{14}) the following expression for the density of the current
\begin{equation}\label{28}
 {\bf j}_l=\frac{\hbar}{m}\rho_l\nabla\varphi_l+\frac{\hbar}{
 m S}\sum_{{\bf k }} {\bf k} \langle b_{l}^+({\bf k}) b_{l}({\bf
 k})\rangle.
\end{equation}
Substituting Eq. (\ref{20}) with coefficients (\ref{22}),
(\ref{23}) into Eq. (\ref{28}), computing the averages and
expanding the result in powers of the phase gradients we obtain
the following expression for the currents:
\begin{equation}\label{29}
  {\bf j}_1=\frac{\hbar}{m}\left[(\rho_{{\rm s}1}-\rho_{{\rm dr}})\nabla\varphi_1
  +\rho_{{\rm dr}}\nabla\varphi_2
  \right],
\end{equation}
\begin{equation}\label{30}
  {\bf j}_2=\frac{\hbar}{m}\left[(\rho_{{\rm s}2}-\rho_{{\rm dr}})\nabla\varphi_2
  +\rho_{{\rm dr}}\nabla\varphi_1
  \right].
\end{equation}
Eqs. (\ref{29}), (\ref{30}) are given in linear in $\nabla
\varphi_l$ approximation. Here the higher order in the phase
gradients terms can be neglected if the phase gradients
$\nabla\varphi_l$ are much smaller than the inverse healing
lengths $\xi^{-1}_l\sim \sqrt{m\gamma\rho_l}/\hbar$ (that
corresponds to the velocities of the superflow much smaller than
the critical ones $v_{lc}\sim\sqrt{\gamma\rho_l/m}$). In Eqs.
(\ref{29}), (\ref{30}) the quantities $\rho_{{\rm s}l}$ and
$\rho_{{\rm dr}}$ with an accuracy up to the $g_k^2$ are
determined by the expressions
\begin{eqnarray}\label{31}
   \rho_{{\rm s}l}=\rho_{l}+\frac{1}{S}\sum_{\bf k} \varepsilon_k
   \frac{\partial N_{lk}}{\partial E_{lk}}-
   \frac{1}{S}\sum_{\bf k} g_k^2 \epsilon_k \Bigg[(-1)^l\left(\frac{\partial N_{1k}}{\partial
   E_{1k}}-\frac{\partial N_{2k}}{\partial
   E_{2k}}\right)\left(\frac{1}{(E_{1k}+E_{2k})^2}-\frac{1}{(E_{1k}-E_{2k})^2}
\right)\cr+\frac{\partial^2 N_{lk}}{\partial
   E_{lk}^2}\left(\frac{1}{E_{1k}+E_{2k}}+\frac{(-1)^l}{E_{1k}-E_{2k}}\right)\Bigg],
\end{eqnarray}
\begin{eqnarray}\label{32}
   \rho_{{\rm dr}}=
   \frac{2}{S}\sum_{\bf k} g_k^2 \epsilon_k \Bigg[
\frac{1+N_{1k}+N_{2k}}{(E_{1k}+E_{2k})^3}-
\frac{N_{1k}-N_{2k}}{(E_{1k}-E_{2k})^3} +\frac{1}{2}
   \left(\frac{\partial N_{1k}}{\partial
   E_{1k}}+\frac{\partial N_{2k}}{\partial
   E_{2k}}\right)\times \cr\left(\frac{1}{(E_{1k}-E_{2k})^2}-
\frac{1}{(E_{1k}+E_{2k})^2}\right)\Bigg].
\end{eqnarray}
Here $N_{l k} =[\exp( E_{l k}/T)-1]^{-1}$  is the Bose
distribution function. One can see that in the absence of the
interlayer interaction ($g_k=0$) the value of $\rho_{{\rm dr}}$ is
equal to zero, and Eq. (\ref{31}) for $\rho_{{\rm s}l}$ is reduced
to the well known expression for the density of the superfluid
component at finite temperatures. If the interlayer interaction is
switched on,  the value of $\rho_{{\rm dr}}$ becomes nonzero.
Then, even in the absence of the phase gradient in the drag layer
the superfluid current in this layer emerges as a response on the
phase gradient in the drive layer.

Eqs.  (\ref{31}), (\ref{32}) were derived under assumption of
$\rho_1\ne \rho_2$ (and, consequently $E_{1k}\ne E_{2k}$). The
case $\rho_1\approx \rho_2$ required more rigorous consideration
since in this case the mixing of the modes is strong even for a
weak interlayer interaction. One can find that the expressions
(\ref{31}), (\ref{32}) remain finite at $\rho_{1}\to \rho_{2}$:
\begin{eqnarray}\label{33}
   \lim_{\rho_1\to \rho_2}\rho_{{\rm s}1}= \lim_{\rho_1\to \rho_2}\rho_{s2}
   =\frac{1}{S}\sum_{\bf k} \varepsilon_k
   \frac{\partial N_{k}}{\partial E_{k}}-
   \frac{1}{2S}\sum_{\bf k} \frac{g_k^2
   \epsilon_k}{E_k} \left(\frac{\partial^2 N_{k}}{\partial
   E_{k}^2}- E_k\frac{\partial^3 N_{k}}{\partial
   E_{k}^3}\right),
\end{eqnarray}
\begin{eqnarray}\label{34}
    \lim_{\rho_1\to \rho_2}\rho_{{\rm dr}}=
   \frac{1}{4S}\sum_{\bf k} \frac{g_k^2 \epsilon_k}{E_k^3}
   \left(1+2
   N_k-2 E_k\frac{\partial N_{k}}{\partial
   E_{k}}+\frac{2}{3} E_k^3\frac{\partial^3 N_{k}}{\partial
   E_{k}^3}\right),
\end{eqnarray}
where $E_k$ is the energy of the elementary excitations at
$\rho_1=\rho_2$ and $\gamma'=0$. Using the exact expressions for
the spectra and the $u-v$ coefficients  we obtain that for the
case of two layers with equal densities and in the weak interlayer
interaction limit $\gamma'\ll \gamma$ the quantities $\rho_{\rm
s}$ and $\rho_{\rm dr}$ are determined just Eqs. (\ref{33}),
(\ref{34}). It allows us to conclude that Eqs. (\ref{31}),
(\ref{32}) are valid for an arbitrary ratio between the densities.

Let us fist consider the case of zero temperature. We define the
drag current as the current in the drag layer (e.g. layer 1) in
the absence of the phase gradient in this layer.  At $T=0$ the
drag current is equal to
\begin{eqnarray}\label{35}
   j_{dr}=C_{dr}\left(\frac{\gamma'}{\gamma}\right)^2 \left(\frac{m \gamma}{2\pi
   \hbar^2}\right)
   \frac{\hbar}{m}\rho_1  \nabla\varphi_2,
   \end{eqnarray}
where
  \begin{eqnarray}\label{36}
    C_{dr}= \int_0^\infty d x
   \frac{x^{1/2} }
   {\sqrt{x+1}\sqrt{x+\rho_1/\rho_2}
   \left(\sqrt{x+1}+\sqrt{x+\rho_1/\rho_2}\right)^3}.
   \end{eqnarray}
The factor $C_{dr}$  is an increasing function of the   ratio
$\rho_2/\rho_1$ (at $\rho_2/\rho_1\to 0$ the factor $C_{dr}$
approaches to zero, at $\rho_2= \rho_1$ it is equal to 1/12, and
it approaches to 1/4 at $\rho_2/\rho_1\to \infty$).  Thus, the
drag current increases under  increasing  the density of the
particles in the drive layer.

\begin{figure}
\begin{center}
\includegraphics[width=16cm]{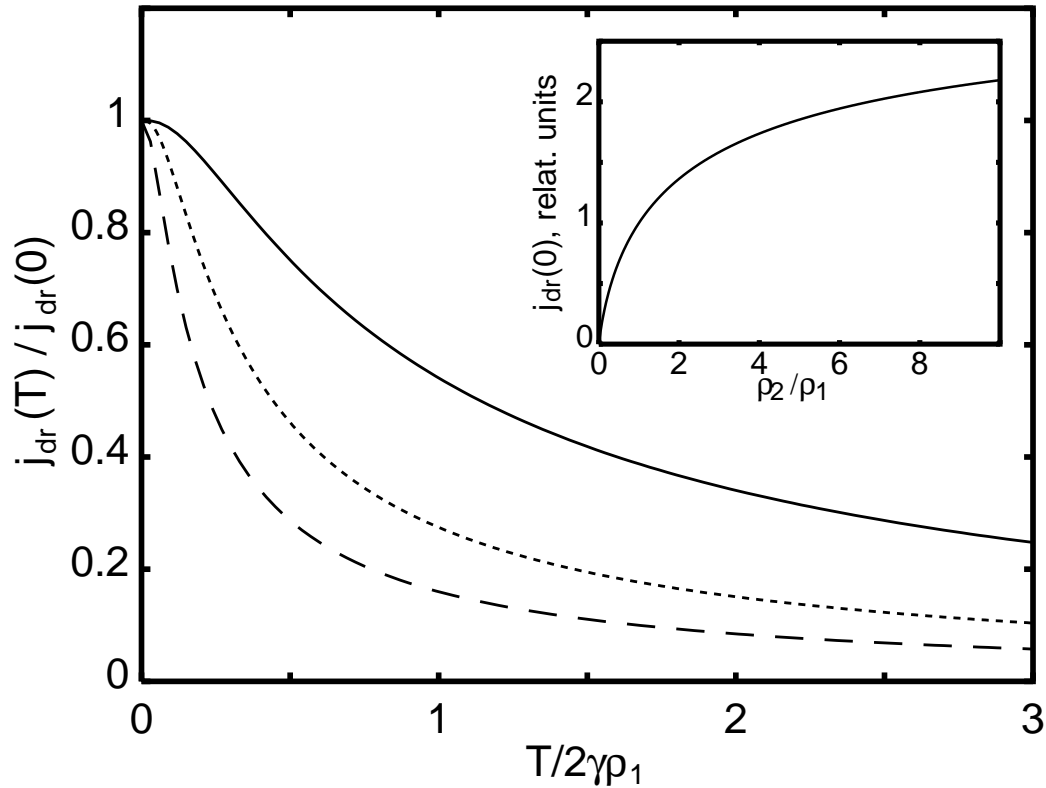}
 \caption{\label{f1} The dependence of the drag current on the
 temperature
 at $\rho_2/\rho_1=5,1,0.2$ (solid, dashed and dotted curves,
 correspondingly) normalized to its value  at $T=0$.
 The dependence of $j_{dr}(0)$ on the ratio $\rho_2/\rho_1$ normalized to its value
 at  $\rho_2=\rho_1$ is shown in the inset.}
 \end{center}
\end{figure}

At finite temperatures the drag current decreases. At small $T$
 one can use the long-wave approximation for the spectra $E_{1(2)k}$  in
the temperature dependent part of Eq. (\ref{32}) and evaluate this
part analytically. It yields the following relation:
\begin{equation}\label{37}
   j_{dr}(T)=
j_{dr}(0)\left[1-\frac{16\zeta(3)}{C_{dr}} \frac{\rho_1}{\rho_2}
  \left(\frac{T}{2\gamma\rho_1}\right)^3 \right].
\end{equation}
But, actually, this approximation is valid only at very low
temperatures.  The results of numerical evaluation of  Eq.
(\ref{32}) are shown in Fig. \ref{f1}. This figure demonstrates
that at $T\sim 2 \gamma \rho_1$ a temperature decrease of the drag
current is much slower.
 Basing on the results presented in Fig. \ref{f1} we also conclude that
 the temperature reduction of the
drag current becomes smaller under increasing  the density of the
particles  in the drive layer.

In a Bose cloud confined in a trap the density is nonuniform. It
results in a modification of the spectrum of elementary
excitations. We may argue that this modification reveals itself in
only minor changes of the value of the drag.  One can  find that
the main contribution to the sum in Eq. (\ref{32}) comes from the
excitations with the wave vectors of order  or higher than the
inverse healing lengths $\xi^{-1}_l$.  In systems with the healing
lengths much smaller than the linear size of the Bose clouds
 the spectrum at $q \gtrsim \xi^{-1}_1,\xi^{-1}_2$ is
well described by the quasi uniform approximation. Therefore, in
such systems the local drag current is given by the same equations
(\ref{29})-(\ref{32}), as in the uniform case with the only
modification that the densities $\rho_1$, $\rho_2$ in these
formulas should be understood as the functions of ${\bf r}$. In
particular, we predict, that for two Bose gases confined in
harmonic traps having the same Thomas-Fermi radius (and this
radius is much larger than the average healing length) a spatial
distribution of the superflow in the drag trap $T=0$ will
replicate (with a drag factor) the spatial distribution of the
superflow in the drive trap. At finite temperatures one can expect
a reduction of the drag factor near the edge of the Bose cloud,
where the density is small.

 One can ask to which extent
two-dimensionality of the system studied may influence on the
results obtained. It is known that in 2D systems fluctuations of
the phase of the order parameter are large  and at nonzero
temperature the off-diagonal one-particle density matrix $\langle
\hat{\Psi}^+({\bf r})\hat{\Psi}(0)\rangle$ goes to zero in the
limit $|{\bf r}|\to\infty$. It means the absence of the long range
order in the systems at $T\ne 0$. But since the asymptotic
behavior of the density matrix are described by a power-law
dependence on $r$ (not an exponential one), at temperatures lower
than the critical one (the Kosterlits-Touless transition
temperature) the superfluid density becomes nonzero. The drag of
the superflow between two 2D Bose-gases, considered in the paper,
is connected with a finite value of the superfluid density and
that is why it decreases under increasing  the temperature. The
density and phase operator approach, used in this paper, is not
based on the existence of the Bose-Einstein condensate.  Moreover,
the power-law asymptotic behavior of the density matrix can be
easily derived in this approach under accounting the thermal
excitations described by the Hamiltonian $H_2$. But at the same
level of approximation we do not find any crucial influence of the
two-dimensionality on the drag phenomena.

\section{The value of the drag in atomic Bose gases}
\label{s4}

Let us present some estimates for the value of the drag in atomic
Bose gases. For simplicity we specify the case of
$\rho_1=\rho_2=\rho$ and $T=0$. It is convenient to introduce the
drag factor $f_{\rm dr}=\rho_{\rm dr}/(\rho-\rho_{\rm dr})$  that
gives the ratio between the currents in the drag and in the drive
traps in the absence of the phase gradient in the drag trap.
Taking into account that $\rho_{\rm dr}\ll\rho$,  we use the
approximate expression $f_{\rm dr}=\rho_{\rm dr}/\rho$ for further
analysis.

 The
value of $f_{dr}$ depends on the interaction parameters $\gamma$
and $\gamma'$. The parameter $\gamma$ can be expressed through the
dimensionless effective "scattering length" $\tilde{a}$
\begin{equation}\label{38}
  \gamma=\frac{2\sqrt{2\pi}\hbar^2}{m} \tilde{a}.
\end{equation}
In  a quasi two-dimensional trap the
 effective scattering length is connected with the 3-D scattering length  $a$ and
the oscillator length in $z$ direction
$l_z=\sqrt{\hbar/m\omega_z}$ by the relation $\tilde{a}= a/l_z$,
which is valid for $a\ll l_z$ \cite{16}. Let us introduce the
interlayer dimensionless effective "scattering length"
$\tilde{a}'$ that is connected with the interlayer interaction
parameter $\gamma'$ by the relation
\begin{equation}\label{39}
  \gamma'=\frac{2\sqrt{2\pi}\hbar^2}{m} \tilde{a}'.
\end{equation}
Substituting Eqs. (\ref{38}), (\ref{39}) into Eq. (\ref{34}) we
obtain the drag factor in the form
\begin{equation}\label{40}
  f_{\rm dr}=\frac{1}{12}\sqrt{\frac{2}{\pi}}\frac{(\tilde{a}')^2}{\tilde{a}}.
\end{equation}
Eq. (\ref{40}) is valid for $|\tilde{a}'|\ll \tilde{a}$, but one
can expect that it is  approximately correct at
$|\tilde{a}'|\approx \tilde{a}$ (we emphasize that in any case the
stability condition requires the inequality $|\tilde{a}'|<
\tilde{a}$ be satisfied).

 To evaluate the value of $|\tilde{a}'|$
we should specify a mechanism of the interlayer interaction. Let
us first consider the interaction that corresponds to the "tail"\
of the Van der Waals potential:
\begin{equation}\label{41}
  V_{12}^{\rm VdW}(r)=-\frac{C_6}{(r^2+d^2)^3}.
\end{equation}
Here $C_6$ is the Van der Waals constant and $d$ is the interlayer
distance. The Fourier-component of the potential (\ref{41}) is
\begin{equation}\label{42}
 {\cal V}^{\rm VdW}_{12}(k)=\int d^2 r
 V_{12}^{\rm VdW}(r)e^{i{\bf k r}}=-\frac{\pi C_6}{4 d^2} k^2
K_2(k d),
\end{equation}
where $K_2(x)$ is the modified Bessel function of the second kind.
Taking into account that the Van der Waals interaction is
 a short-ranged one we can evaluate $\gamma'$ as $\gamma'={\cal V}^{\rm
VdW}_{12}(k\to 0)=-\pi C_6/(2 d^4)$. It yields
$|\tilde{a}'_{VdW}|\approx \sqrt{\pi/2}C_6 m/(4\hbar^2 d^4)$.

\begin{figure}
\begin{center}
\includegraphics[width=10cm]{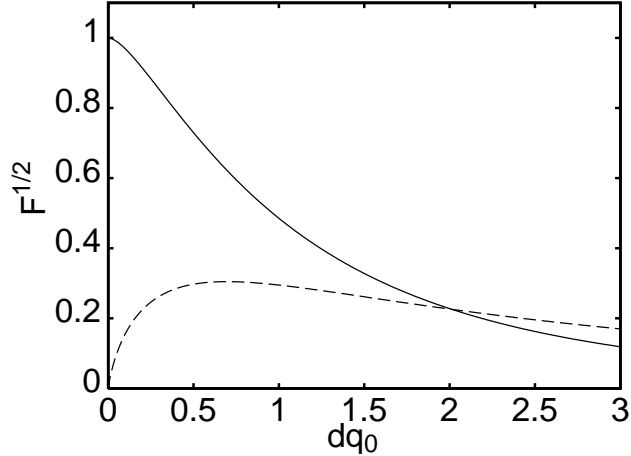}
 \caption{\label{f2} The density dependent factors in the effective
  interlayer scattering length
versus  the parameter $dq_0=\sqrt{8\pi\rho d^2/\tilde{a}}$. Solid
curve - $\sqrt{F_{VdW}}$ (Van der Waals interaction), dashed curve
-  $\sqrt{F_{d-d}}$ (dipole-dipole interaction).}
 \end{center}
\end{figure}

This result can be obtained in a more rigorous way. Our approach
is easily generalized for the case of an arbitrary central force
interlayer interaction potential. To do this we should redefine
the quantity $g_k$ as
\begin{equation}\label{43}
  g_k= {\cal V}_{12}(k) \epsilon_k
  \sqrt{\frac{\rho_1\rho_2}{E_{1k}E_{2k}}}.
\end{equation}
and substitute this definition (instead of Eq. (\ref{19})) into
the formulas for $\rho_{{\rm s} l}$ and $\rho_{\rm dr}$ obtained
in the previous section. Using Eq. (\ref{34}), one can present the
drag factor (for $T=0$ and $\rho_1=\rho_2$) in the following form
\begin{equation}\label{44}
f_{\rm dr}=\frac{1}{16
\pi^2}\frac{m^2}{\hbar^4 \tilde{a}}
\sqrt{\frac{\pi}{2}}\int_0^\infty d x \frac{x^2 [{\cal V}_{12}(q_0
x)]^2}{(x^2+1)^{5/2}}
\end{equation}
with $q_0=\sqrt{8\pi \rho \tilde{a}}$. Substituting  Eq.
(\ref{42}) into Eq. (\ref{44}) we find
\begin{equation}\label{45}
  f_{\rm dr}=\frac{1}{12}\left(\frac{C_6 m}{4\hbar^2
  d^4}\right)^2\frac{1}{\tilde{a}} \sqrt{\frac{\pi}{2}}F_{\rm VdW}(d q_0 ).
\end{equation}
Here the function
\begin{equation}\label{46}
  F_{\rm VdW}(x)=\frac{3x^4}{4}\int_0^\infty d y
  \frac{y^6}{(1+y^2)^{5/2}}K_2^2(x
  y)
\end{equation}
describes the dependence of the drag factor on the density $\rho$.
Comparing Eqs. (\ref{46}) and (\ref{40}) we obtain the following
expression for the modules of the effective interlayer scattering
length
\begin{equation}\label{47}
|\tilde{a}'_{\rm VdW}|= \sqrt{\frac{\pi}{2}}\frac{C_6 m}{4\hbar^2
  d^4}\sqrt{F_{VdW}(dq_0)}.
\end{equation}
The dependence of the factor $\sqrt{F_{VdW}}$  on the parameter $d
q_0$ is shown in Fig. \ref{f2}. One can see from this figure that
at $dq_0\ll 1$ (that corresponds to the low density limit) the
factor $\sqrt{F_{VdW}}$ in Eq. (\ref{47}) is close to unity and we
arrive to the expression for $|\tilde{a}'_{\rm W dV}|$ given
above.

Due to a short-range nature of the Van der Waals interaction the
interlayer effective scattering length decreases quickly under
increasing of $d$. Therefore, the interlayer distance $d$ should
be rather small to achieve an observable value of the drag. Using
a typical value of $C_6$ ($C_6\approx 3\cdot 10^{-57}$
erg$\cdot$cm$^6$) and taking $d\approx 10$nm , $m=87$ a.u. (Rb) we
evaluate $|\tilde{a}'_{\rm VdW}|\approx 10^{-1}$.

The quantities $l_z$ and $a$ can be controlled in experiments. The
first one is controlled by changing the profile of the confining
potential in $z$ direction, and the latter one - by tuning the
magnetic field to the value close to the Feshbach resonance field
 \cite{17}-\cite{19}.
 Near this resonance the scattering length change its sign and a
situation with rather small 3D scattering length $a$ (much smaller
than the value of $l_z$ which, in its turn, has to be smaller than
$d/2$) can be realized. Using this possibility one can tune the
quantity $\tilde{a}$ close to the value of $|\tilde{a}'_{\rm
VdW}|$ and obtain the drag factor $f_{dr}\approx 7\cdot 10^{-3}$.

Out of the resonance the typical values of 3D scattering length
lie in the interval  $3\div 5$nm and for $l_z<d/2$ and $d\approx
10$nm the estimation $\tilde{a}= a/l_z$ is not applicable. In the
ultra 2D limit $(l_z/a\ll 1$) the interaction parameter can be
evaluated by using the formula \cite{20} $$ \gamma =\frac{4 \pi
\hbar^2} {m} \frac{1}{|\ln(\rho a^2)|}.$$ For typical densities
$\rho=10^8\div 10^{10}$ cm$^{-2}$ it yields $\tilde{a}=0.2\div
0.4$ and the drag factor $f_{dr}\approx 2\div 3\cdot 10^{-3}$.

At $d\gtrsim 100$nm the drag caused by the Van der Waals
interaction becomes negligible small. But in the last case the
dipole-dipole interaction may give an essential contribution to
the drag. Let us consider the situation where the  dipole momenta
of the atoms are aligned in a direction perpendicular to the
layers. Then the interaction potential has the form
\begin{equation}\label{48}
  V_{12}^{d-d}(r)=D^2 \frac{r^2-2 d^2}{(r^2+d^2)^{5/2}},
\end{equation}
where $D$ is the dipole momentum.  The Fourier-component of the
potential (\ref{48}) reads as
\begin{equation}\label{49}
 {\cal V}_{12}^{d-d}(k)=-2 \pi D^2 k e^{-k d}.
\end{equation}
Substituting  Eq. (\ref{49}) into Eq. (\ref{44})  we obtain
\begin{equation}\label{50}
  f_{\rm dr}=\frac{1}{12}\left(\frac{D^2 m}{\hbar^2
  d}\right)^2\frac{1}{\tilde{a}} \sqrt{\frac{\pi}{2}}F_{d-d}(d q_0),
\end{equation}
where
\begin{equation}\label{51}
  F_{d-d}(x)=3 x^2\int_0^\infty d y
  \frac{y^4}{(1+y^2)^{5/2}}e^{-2 x
  y}
\end{equation}
One can see that Eq. (\ref{50}) is reduced to Eq. (\ref{40}) under
definition
\begin{equation}\label{52}
|\tilde{a}'_{d-d}|= \frac{D^2 m}{\hbar^2
  d}\sqrt{\frac{\pi}{2}}\sqrt{F_{d-d}(dq_0)}.
\end{equation}
The dependence $\sqrt{F_{d-d}(dq_0)}$ is also shown in Fig.
\ref{f2}. In difference with the previous case the value of
$\tilde{a}'_{d-d}$ approaches to zero in the low density limit.
But at $dq_0> 0.1$ that corresponds to $\rho> 10^{-2}
d^{-2}/(8\pi\tilde{a})$ one can neglect the dependence of $f_{\rm
dr}$ on the density and put the factor $\sqrt{F_{d-d}}\approx
0.2$. For the estimates given below we assume the condition
$dq_0>0.1$ is fulfilled.

For the magnetic dipole-dipole interaction $D$ is the magnetic
dipole momentum of the atoms. The magnetic dipoles can be aligned
in the same direction if a constant magnetic field is applied to
the system. Taking $d=100$nm,  $D=\mu_B$ (the Bohr magneton) and
$m$=87 a.u. we obtain $|\tilde{a}'_{d-d}|\approx 3\cdot10^{-4}$.
In the case, tuning $\tilde{a}$ to the $|\tilde{a}'_{d-d}|$ value,
one can achieve the drag factor $f_{dr}\approx  2\cdot10^{-5}$

For the Bose atoms with large magnetic dipole momenta this value
can be much larger.  A good candidate atom is Cr ($D=6 \mu_B$).
The possibility to realize Cr Bose-Einstein condensate is
discussed in \cite{21}. For $m$=52 a.u., $D=6 \mu_B$ and $d=100$nm
we evaluate $|\tilde{a}'_{d-d}|\approx 6\cdot10^{-3}$ and,
consequently, the maximum drag factor $f_{dr}\approx
4\cdot10^{-4}$

\section{The "drag force" as an analog of the vector potential}
\label{s5}

%\vskip 0.5cm

In  section \ref{s3} we compute the drag current directly. The
same results can be obtained from the analysis of the dependence
of the free energy of the system on the phase gradients. The free
energy of the system can be found from the common thermodynamic
relation
\begin{equation}\label{53}
  F=H_0 +E_{zero}+ T \sum_{\lambda=\alpha,\beta}\sum_{{\bf k}}
   \ln\left[1-\exp\left(-\frac{{\cal
  E}_{\lambda}( {\bf k})}{T}\right) \right].
 \end{equation}
Here the quantity $H_0$ given by Eq. (\ref{7}) is the classical
energy of the system, and
\begin{equation}\label{54}
  E_{zero}=\frac{1}{2}\sum_{\lambda=\alpha,\beta}\sum_{{\bf k}}
  \left[{\cal E}_{\lambda}( {\bf k})
   - \epsilon_k \right]
\end{equation}
is the energy of the zero-point fluctuations.

Substituting the spectra (\ref{26}), (\ref{27}) into Eq.
(\ref{53}) and expanding the final expression in powers of the
phase gradients we find the following expression for the free
energy
\begin{equation}\label{55}
  F= F_0+ \int d^2 r \frac{\hbar^2}{2 m}\left[\rho_{{\rm s}1}(\nabla\varphi_1)^2
  +\rho_{{\rm s}2}(\nabla\varphi_2)^2 -\rho_{\rm dr}
  (\nabla\varphi_1-\nabla\varphi_2)^2\right]+ \textrm{higher order
  terms},
\end{equation}
where $F_0$ does not depend on the phase gradients and the
quantities $\rho_{{\rm s}l}$, $\rho_{\rm dr}$ are determined by
the expressions (\ref{31}),(\ref{32}). One can see that the answer
(\ref{29})-(\ref{32}) obtained in Sec. \ref{s3} by another method
can also be found from Eq. (\ref{55}) using the relation
\begin{equation}\label{56}
  {\bf j}_l=\frac{1}{\hbar S}\frac{\partial  F}{\partial
  (\nabla\varphi_l )}.
\end{equation}

The relation (\ref{55}) is more instructive in a sense that it
demonstrates an analogy between the drag effect in superfluids and
the exciting of a supercurrent by an external magnetic field in
superconductors. To illustrate this analogy let us consider two
ring-shape traps and fix the phase gradient in the drive trap
(trap 2 in further notations). Then the free energy as the
function of the phase gradient in the drag trap (trap 1) can be
presented in the form
\begin{equation}\label{57}
F= const+\frac{\pi\hbar^2 w}{m R} \tilde{\rho}_{{\rm s}1}
(\Phi+\Phi_{\rm dr})^2,
\end{equation}
where $R$ is the radius of the ring, $w$, its width,
$\tilde{\rho}_{{\rm s}1}=\rho_{{\rm s}1}-\rho_{\rm dr}$,
\begin{equation}\label{58}
  \Phi=\frac{1}{2 \pi}\oint_C d {\bf l} \nabla\varphi_1
\end{equation}
(here  $C$ is a contour around the ring) is the winding number for
the phase $\varphi_1$ and
\begin{equation}\label{59}
\Phi_{\rm dr}=\frac{\rho_{\rm dr}}{\tilde{\rho}_{{\rm s}1}}
\frac{1}{2 \pi}\oint_{C} d {\bf l} \nabla\varphi_2,
\end{equation}
is the winding number for the phase $\varphi_2$ times the drag
factor. In deriving (\ref{57}) we, for simplicity, neglect the
dependence of densities on the coordinate inside the traps.

Since the value of $\Phi$ should be integer the minimum of the
free energy at $|\Phi_{\rm dr}|<1/2$ is reached for $\Phi=0$. In
this case  the phase gradient in the drag trap is equal to zero
and the superfluid current in the drag trap flows in  the same
direction as in the drive trap. If $|\Phi_{\rm dr}|>1/2$  the free
energy reaches its minimum at nonzero $\Phi$ and the phase
gradient is induced in the drag trap. Then together with the drag
current the counterflow current appears in the drag trap
(depending on the value of $\Phi_{\rm dr}$ the total current in
this trap can be parallel as well as antiparallel to the current
in the drag trap). Just the same situation takes place in a
superconducting ring with nonzero flux of magnetic field inside
the ring. Thus, in  two-ring Bose systems the quantity $\Phi_{\rm
dr}$ plays the same role as a flux of an external magnetic field
(measured in units of flux quanta) in superconducting circuits.

To realize this situation experimentally one should create a
circulating superflow in the drive trap. It can be done by
elliptic rotating deformation of this trap. The rotation can be
switched of when a superflow be created. The value of drag current
can be found from measurement of the angular momentum of the drag
trap. At present a number methods for measuring this quantity has
been realized experimentally \cite{22}-\cite{25}. The methods are
based on the study of dynamics of collective excitations, on the
investigation of interference phemonena under hyperfine state
transitions and on the observation of the dynamics of expansion of
the Bose cloud.

To extend the analogy with superconductors let us consider the
case where the drag trap of a ring geometry contains a weak
Josephson link. Then the free energy as the function of the phase
shift $\Delta\varphi$ on the link reads as
\begin{equation}\label{60}
  F= const -E_J \cos(\Delta\varphi)+ \frac{\pi\hbar^2 w}{m R} \tilde{\rho}_{{\rm s}1}
\left(\frac{\Delta\varphi}{2\pi}+\Phi_{\rm dr}\right)^2,
\end{equation}
where $E_J$ is the Josephson energy. At $E_J>(E_J)_{c}=  \hbar^2 w
\tilde{\rho}_{{\rm s}1} /(2\pi R m)$ and $|\Phi_{\rm dr}|=
1/2,3/2,\ldots$ the dependence $F(\Delta\varphi)$ has two
degenerate minima. If $E_J/(E_J)_{c}-1\ll 1$  these minima are
very shallow and one can expect that two quantum states with
different phase shifts (and with the superfluid currents flowing
in opposite directions) will be entangled. It is the same regime
that is required for implementing the superconducting Josephson
flux (persistent current) qubit \cite{26}. While in alkali-metal
Bose gases the drag factor is rather small and  even in the most
favorable conditions the maximum value can be reached is of order
$10^{-2}\div 10^{-3}$ (see Sec. \ref{s4}), the case $|\Phi_{\rm
dr}|\approx 1/2$ can be realized in ring-shape traps of large
radiuses ($10^2\div 10^3$ $\mu$m).

In conclusion, we would like to mention another systems in which
the effects described in this paper may take place. It is
excitonic or electron-hole Bose liquids in electron bilayers. In
these systems electron-hole pairs with components belonging to
adjacent layers may form a superfluid state. For the first time
the effect was predicted in \cite{27,28}, and recently it was
confirmed experimentally \cite{29}. The superfluid drag effect may
emerge in two parallel bilayers (the four-layer system). In the
four-layer system the intralayer (in the same bilayer) and
interlayer (between bilayers) interactions are of the same order:
both of them are determined by the dipole-dipole mechanism. In
such a case the dipole momentum of the pair is large. Therefore,
one can expect that nondissipative drag in these systems will be
rather strong.

This work is supported by the INTAS grant No 01-2344.

\end{document}